\documentclass[twocolumn,showpacs,preprintnumbers,amsmath,
amssymb,superscriptaddress,aps,pra]{revtex4}
\usepackage{graphicx}% Include figure files
\usepackage{bm}% bold math
\begin{document}
\preprint{APS/123-QED}
\title{FDTD Simulation of Thermal Noise in Open Cavities}
\author{Jonathan Andreasen}
\author{Hui Cao}
\email{h-cao@northwestern.edu}
\affiliation{
  Department of Physics and Astronomy, Northwestern University,
  Evanston, Illinois 60208-3112
}
\author{Allen Taflove}
\affiliation{
  Department of Electrical Engineering and Computer Science, 
  Northwestern University,
  Evanston, Illinois 60208-3112
}
\author{Prem Kumar}
\affiliation{
  Department of Physics and Astronomy, Northwestern University,
  Evanston, Illinois 60208-3112
}
\affiliation{
  Department of Electrical Engineering and Computer Science, 
  Northwestern University,
  Evanston, Illinois 60208-3112
}
\author{Chang-qi Cao}
\affiliation{
  Department of Physics, Peking University,
  Beijing 100871, China
}
\date{\today}
\begin{abstract}
  A numerical model based on the finite-difference time-domain (FDTD) method
  is developed to simulate thermal noise in open cavities owing to output 
  coupling. 
  The absorbing boundary of the FDTD grid is treated as a blackbody, 
  whose thermal radiation penetrates the cavity in the grid. 
  The calculated amount of thermal noise in a one-dimensional dielectric 
  cavity recovers the standard result of the quantum Langevin equation in 
  the Markovian regime. 
  Our FDTD simulation also demonstrates that in the non-Markovian regime the 
  buildup of the intracavity noise field depends on the ratio of the cavity 
  photon lifetime to the coherence time of thermal radiation.
  The advantage of our numerical method is that the thermal noise is 
  introduced in the time domain without prior knowledge of cavity modes. 
\end{abstract}
%05.40.-a Fluctuation phenomena, random processes, noise, and Brownian motion
%42.25.Kb Coherence
%44.40.+a Thermal radiation
\maketitle 

\section{Introduction}

The finite-difference time-domain (FDTD) method \cite{tafl05} has been 
extensively used in solving Maxwell's equations for dynamic electromagnetic 
(EM) fields. 
The absorbing boundary condition based on the perfectly matched layer (PML) 
\cite{berenger94} allows the simulation of open systems, e.g. leaky optical 
cavities, in any dimension. 
The incorporation of auxiliary differential equations, such as the rate 
equations for atomic populations \cite{nagra98} and the Maxwell-Bloch 
equations for the density-of-states of atoms \cite{ziol95,ziol97AO,ziol97IEEE}, 
has lead to comprehensive studies of light-matter interactions. Although the 
FDTD method has become a powerful tool in computational electrodynamics, 
it is  applied mostly to classical or semiclassical problems. 
Recently, quantum fluctuations due to the spontaneous emission of atoms were 
introduced to the FDTD simulation of microcavity lasers \cite{hof99,slav04}. 
The light field in an open cavity also experiences quantum fluctuation 
because of its coupling to external reservoirs. 
In this paper, we model the quantum noise for the cavity field as a classical 
noise and incorporate it into the FDTD algorithm.

There are two dissipation mechanisms for the cavity field: (i) intracavity 
absorption, (ii) output coupling. 
In the modal picture, widely used in quantum optical studies, 
thermal noise is introduced so that the quantum operator of a leaky cavity 
mode satisfies the commutation relation. 
Although thermal noise is quantitatively insignificant at optical 
frequencies, its proper treatment constitutes an essential part of the exact 
quantum-mechanical theory of lasers. 
Early laser theory introduces the thermal noise via a heatbath made up of 
loss oscillators or absorbing atoms \cite{haken83,lax}. 
It accounts for light absorption inside the cavity. 
For a laser cavity whose loss only comes from the output coupling, 
the thermal noise is attributed to the thermal radiation that penetrates 
the cavity through the coupling \cite{lang73,ujihara77}. 
Thus the amount of thermal noise depends on the mode decay rate, 
which must be known in order to solve the Langevin equation for the field 
operator. 
For open complex cavities, e.g. the ones made of random structures, 
the required information of modes is unknown \textit{a priori}. 
Thus, it is desirable to be able to study the noise of a cavity field without 
prior knowledge of cavity modes. 
Additional problems with the modal picture are, (i)  if the cavity is very 
leaky, the significant overlap of modes in frequency makes it difficult to 
distinguish one mode from another; (ii) In the presence of nonlinearity, 
strictly speaking, the modes do not exist.  
In fact, one advantage of the FDTD method is the direct time-domain 
calculation of EM fields without prior knowledge of modes.  
The effective modal behavior is an \textit{emergent} property that results 
from temporal evaluation of the EM fields. 
We intend to introduce noise to the EM field in a way compatible with the
FDTD method, namely, without invoking the modal picture.
Our goal is to open a new approach for the study of quantum mechanical aspects
of radiation in macroscopic systems with classical electrodynamics simulations.
We believe our approach has the potential to permit rigorous theoretical
investigations of noise in the area of quantum optics and of open systems such
as chaotic open cavities.
The dynamics of such systems are, in particular, very difficult to study using
the standard frequency domain methods.

In FDTD simulations, it is rather straightforward to introduce noise 
related to intracavity absorption. 
A fluctuating electric field can be added as a soft source at every grid 
point inside the cavity with its rms amplitude proportional to the local 
absorption coefficient \cite{luoPRL04,chanPRE06}. 
The output coupling, however, is not a local loss. 
The question is how to introduce thermal noise related to cavity leakage 
without knowing the leakage rate. 
In FDTD simulations, light escaping from an open system is absorbed by the 
absorbing boundary layer (ABL) which acts as the external reservoir. 
Since it absorbs all impinging fields, the ABL can be modeled as a blackbody. 
To remain in thermal equilibrium, the blackbody must radiate into the system. 
The blackbody radiation from the ABL propagates into the cavity and acts as 
noise to the cavity field. 
The amount of noise penetrating the cavity depends on the cavity openness or 
output coupling. 
We simulate the blackbody radiation from the ABL in the FDTD calculation. 
Our model is validated in the calculation of field noise in a one-dimensional 
dielectric cavity. 
In a good cavity whose lifetime $\tau$ is much longer than the coherence time 
of thermal radiation $\tau_c$, the average amount of thermal noise in one 
cavity mode agrees to the solution of the quantum Langevin equation under the 
Markovian approximation. 
In addition to recovering the standard results, our simulations with various 
values of $\tau$ and $\tau_c$ illustrate the transition from the Markovian 
regime to the non-Markovian regime, and demonstrate that the buildup of the 
intracavity noise field depends on the ratio of $\tau_c$ to $\tau$. 
This result is explained qualitatively by interference effect. 

The paper is organized as follows.
Section~\ref{sec:num_method} outlines our numerical method. 
Possible numerical difficulties and problems are discussed.
In Section~\ref{sec:results} we present the results of the FDTD calculations, 
including the blackbody radiation in vacuum and noise penetration into a 
one-dimensional (1D) cavity.
The transition from the Markovian regime to the non-Markovian regime is studied.
Section~\ref{sec:discussion} consists of a discussion and interpretation of 
the results. 
An analytical expression is found which offers further insights to the amount 
of noise inside an open cavity.
We end in Section~\ref{sec:conclusion} by summarizing all results and
including a discussion of future applications for our method.

\section{Numerical Method\label{sec:num_method}}

\begin{figure}
  \includegraphics[width=8.6cm]{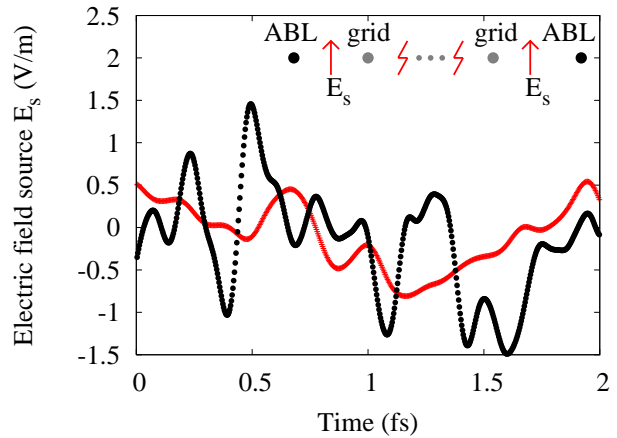}
  \caption{\label{fig:fig1} (Color online) 
    Noise source electric field $E_s(t_j)$ generated 
    for $T=30,000$ K (black dots) and $T=50,000$ K (red crosses).
    The noise correlation time $\tau_c\approx 0.337$ fs for $T=30,000$ K and 
    $\tau_c\approx 0.203$ fs for $T=50,000$ K.
    $\Delta x = 1$ nm, $M=2^{21}$ and $\tau_{sim}=7$ ps.
    The inset is a schematic showing the noise sources placed next to the 
    grid/ABL interface.
  }
\end{figure}

\begin{figure}
  \includegraphics[width=8.6cm]{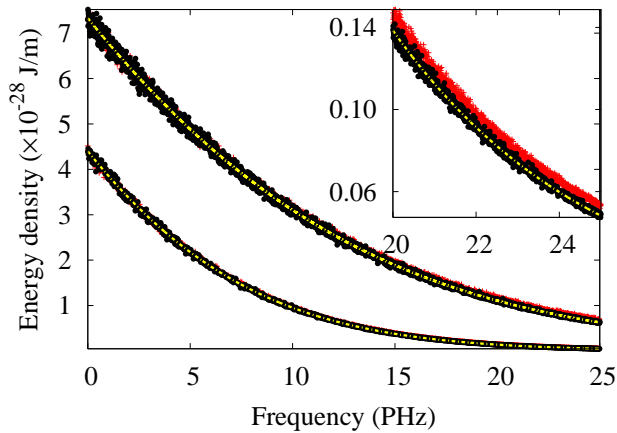}
  \caption{\label{fig:fig2} (Color online) 
    FDTD-calculated energy density of blackbody 
    radiation propagating in 1D vacuum versus frequency $\omega$ for 
    temperatures $T=30,000$ K (lower) and $T=50,000$ K (upper).
    The inset shows the energy density for temperature $T=30,000$ K at higher 
    frequencies.
    The data are obtained by averaging over 2000 calculations with the 
    resolutions $\Delta x=10$ nm (red crosses) and $\Delta x =1$ nm 
    (black dots).
    The source spectra $D(\omega,T)$ are also plotted as solid lines on top 
    of the numerical spectra.
  }
\end{figure}

Our numerical model is based on the key insight that the ABL normally used to 
bound FDTD computational grids is in effect a blackbody which ideally absorbs 
all incident radiation. 
To stay in thermal equilibrium with temperature $T$, the blackbody must 
radiate into the system. 
To simulate the blackbody radiation, we surround the grid with a series of 
noise sources next to the grid/ABL interface. 
These soft sources radiate EM waves into the grid having spectral properties 
consistent with blackbody radiation. 
In this paper, we focus on 1D systems. 
The 1D grid is discretized with a spatial step $\Delta x$ and time step 
$\Delta t$. As shown in the inset of Fig. \ref{fig:fig1},
two point sources are placed at the extremities of the grid. 
Each source generates an electric field $E_s$ at every time step $t_j$.
Examples of the noise source of electric field $E_s(t_j)$ are shown in 
Fig.~\ref{fig:fig1}.
A Fourier transform of the temporal correlation function of the electric 
field, $\langle E_s(t_1) E_s(t_2) \rangle$, gives the noise spectrum 
$D(\omega,T)$.
If $E_s(t_j)$ is uncorrelated in time, i.e., 
$\langle E_s(t_1) E_s(t_2) \rangle \propto \delta (t_2 - t_1)$, $D(\omega,T)$ 
is the white-noise spectrum. 
This is incorrect as $D(\omega,T)$ should be equal to the energy density of 
the blackbody radiation \cite{siegbook,Garrod95}, which in one dimension is
\begin{equation}
  D(\omega,T) = \frac{\hbar}{\pi c}\left(\frac{\omega}
  {\exp(\hbar\omega/kT)-1}\right).\label{eq:bbu}
\end{equation}
$D(\omega,T)$ for two different temperatures is plotted in 
Fig.~\ref{fig:fig2}. 
For computational convenience, we extend the range of $\omega$ from 
$(0, \infty)$ to $(-\infty, \infty)$. 
Since the electric field in the FDTD simulation is a real number, 
$D(-\omega,T)$ must be equal to $D(\omega,T)$ for $\omega > 0$. 
Therefore, $D(\omega,T)=D(|\omega|,T)$.
We normalize $D(\omega,T)$ as
\begin{equation}
  D_n(|\omega|,T) = \frac{6\hbar^2}{\pi k^2T^2}\left(\frac{|\omega|}
    {\exp(\hbar |\omega| / k T)-1}\right)\label{eq:bbn}
\end{equation}
so that $\int_{-\infty}^{\infty} D_n(|\omega|,T) d\omega=2\pi$.

\begin{figure}
  \includegraphics[width=8.6cm]{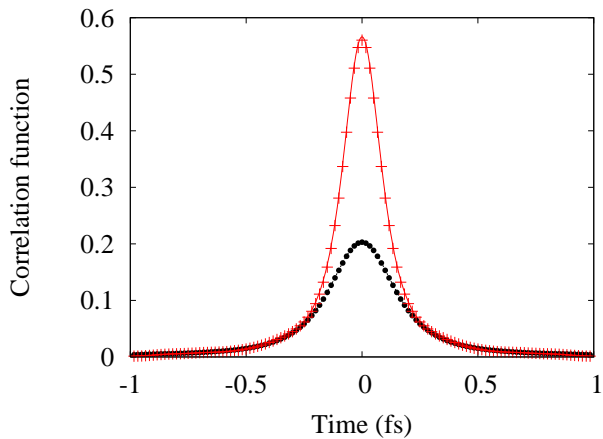}
  \caption{\label{fig:fig3}  (Color online) Temporal correlation function,
    $\left< E_s(t_1)E_s(t_2)\right>$ vs. $t_2-t_1$, for the noise electric
    field at $T=30,000$ K (black circles) and $T=50,000$ K (red crosses).
    The noise correlation times are $\tau_c\approx 0.337$ fs for $T=30,000$ K
    and $\tau_c\approx 0.203$ fs for $T=50,000$ K.
    $\Delta x = 1$ nm, $M=2^{21}$, and $\tau_{sim}=7$ ps.
    The lines represent $\left< E_s(t_1)E_s(t_2)\right>$ given by the
    analytical expression in Eq.~\ref{eq:coran} for $T=30,000$ K (black) and
    $T=50,000$ K (red).  
    Every 5th data point is taken from the numerical data in order to better 
    show the agreement with the analytical solution.
  }
\end{figure}

The temporal correlation function for the source electric field is given by
\begin{equation}
  \left< E_s(t_1) E_s(t_2) \right> = \frac{\delta^2}{2\pi} 
  \int_{- \infty}^{\infty}d\omega D_n(|\omega|,T) e^{i \omega (t_2-t_1)},
  \label{eq:bbcorr}
\end{equation}
where $\delta$ is the rms amplitude of the noise field whose value is to be 
determined later.
For the thermal noise, the field correlation function is given specifically by 
\begin{align}
  \left< E_s(t_1) E_s(t_2) \right> = \frac{3\delta^2}{\pi^2}[
  &\zeta(2,1-i(t_2-t_1)kT/\hbar)\nonumber\\
  +&\zeta(2,1+i(t_2-t_1)kT/\hbar)]
  \label{eq:coran},
\end{align}
where the $\zeta$-function is given as
\begin{equation}
  \zeta (s,a) = \sum_{k=0}^{\infty}(k+a)^{-s}.
\end{equation}
The temporal correlation function of thermal radiation is plotted in 
Fig.~\ref{fig:fig3}.

We employ a quick and straightforward way of generating random numbers for 
$E_s(t_j)$ so that Eq.~\ref{eq:bbcorr} is satisfied.
Freilikher \textit{et al.} have developed such a method in the context of 
creating random surfaces with specific height correlations \cite{mara}.
The end result takes advantage of the fast Fourier transform (FFT) which we 
use to generate the source electric field:
\begin{equation}
  E_s(t_j) = \frac{\delta}{\sqrt{\tau_{sim}}}\sum_{l=-M}^{M-1}(M_l+iN_l)
  D_n^{1/2}(|\omega_l|,T)e^{i\omega_lt_j},\label{eq:noise}
\end{equation}
where $2M$ is the total number of time steps, $\tau_{sim}=2M\Delta t$ is the 
total simulation time and $\omega_l=2\pi l/\tau_{sim}$.
$M_l$ and $N_l$ are independent Gaussian random numbers with zero mean and a 
variance of one.  
Their symmetry properties are $M_l=M_{-l}$ and $N_l=-N_{-l}$.
These Gaussian random numbers can be generated by the Marsaglia and Bray 
modification of the Box-M\"{u}ller Transformation \cite{brys91},
a very fast and reliable method assuming the uniformly distributed random 
number generator is quick and robust.

The electric field sources generate both electric and magnetic fields, which 
propagate into the grid. 
$E(x, \omega)$ and $H(x, \omega)$ are obtained by the discrete Fourier 
transform (DFT) of $E(x,t)$ and $H(x,t)$. 
Since both $E(x,t)$ and $H(x,t)$ are real numbers, 
$E(x, \omega) = E(x, -\omega)$ and $H(x, \omega) = H(x, -\omega)$. 
The EM energy density at frequency $\omega$ shall include $E(x, \omega)$, 
$E(x, -\omega)$, $H(x, \omega)$ and $H(x, -\omega)$. 
If the grid is vacuum, the steady-state energy density at every position $x$ 
should be equal to the blackbody radiation density. 
The rms amplitude $\delta$ of the source field $E_s$ is determined by  
\begin{equation}
  \frac{1}{2}\epsilon_0|E(x, |\omega|)|^2 + \frac{1}{2}\mu_0|H(x,|\omega|)|^2 
  =\frac{\hbar}{\pi c}\frac{|\omega|}{e^{\hbar|\omega|/kT}-1}.\label{eq:FFTEME}
\end{equation}

When setting the parameters in the FDTD simulation, we must taken into 
consideration the characteristics of thermal noise. 
The temporal correlation time or coherence time $\tau_c$ of thermal noise is 
defined as the full width at half maximum (FWHM) of the temporal field 
correlation function. 
If the time step $\Delta t$ is close to $\tau_c$, $E_s$ exhibits a sudden 
jump at each time step. 
The 1D FDTD algorithm cannot accurately propagate such step-like pulses (with 
sharp rising edge) if the Courant factor $S \equiv c\Delta t/\Delta x$ is set 
at a typical value $S<1$. 
The pulse shape is distorted with fringes corresponding to both retarded 
propagation and superluminal response \cite{tafl05}. 
This occurs because the higher frequencies from the step discontinuity are 
being inadequately sampled and because of numerical dispersion arising from 
the method of obtaining the spatial derivatives for $E$ and $H$.
To avoid such problems, we use $S=1$ which eliminates the numerical 
dispersion artifact \cite{tafl05}.
Furthermore, we set $\Delta t \ll \tau_c$ which provides a dense temporal
sampling relative to the correlation/coherence time of the thermal noise.

To obtain an accurate noise spectrum with the DFT, both the frequency and 
temporal resolutions must be chosen carefully. 
The two problems affecting the reliability of the DFT are aliasing and 
leakage due to the use of a finite simulation time \cite{hamming86}. 
The solution to these problems is to increase the number of time steps $2M$ 
and decrease the time step value $\Delta t$. 
This takes the DFT closer to a perfect analytical Fourier transform, but 
run-time and memory limitations must be considered as well. 
Taking advantage of the FFT algorithm significantly reduces both noise 
generation time and spectral analysis time. 

Although the thermal noise spectrum can be very broad, only noise within a 
certain frequency range is relevant to a specific problem. 
Let $\omega_{min}$ and $\omega_{max}$ denote the lower and upper limits of 
the frequency range of interest, and $\Delta\omega$ the frequency resolution 
needed within this range. 
To guarantee the accuracy of the noise simulation in 
$\omega_{min} < \omega < \omega_{max}$, the total running time $\tau_{sim}$ 
must exceed $2\pi/\omega_{min}$ and $2\pi/\Delta\omega$. 
The time step $\Delta t$ has an additional requirement, 
$\Delta t < \pi/\omega_{max}$.

\section{Simulation Results\label{sec:results}}

\subsection{Blackbody radiation in vacuum\label{subsec:vacuum}}

We first test the noise sources in a 1D FDTD system composed entirely of
vacuum. 
Two sets of independent noise signals $E_s(t_j)$ are generated via 
Eq.~\ref{eq:noise}.
One set is added as a soft source one grid cell away from the left absorbing 
boundary; 
the other one cell from the right absorbing boundary.
Both have equal rms amplitude $\delta$ so that the average EM flux to the 
left equals that to the right at any position $x$ in the grid. 
Since the system is one dimensional, the EM flux at any distance away 
from the source has the same magnitude. 
The value of $\delta$ shall be adjusted so that Eq.~\ref{eq:FFTEME} is 
satisfied.
For the EM energy density radiated by one source to equal $D(|\omega|,T)$, 
we set $\delta$ to
\begin{equation}
  \delta = \sqrt{\frac{2}{\epsilon_0}\frac{1}{6 \hbar c}}kT\label{eq:delta}.
\end{equation}
After the noise fields in the grid reach steady state, the noise spectrum 
at any grid point is obtained by a DFT. 
We verify that the spectrum of EM energy density at any point in the grid is 
identical to that at the source. 
This means there is no distortion of the noise spectrum by the propagation of 
noise fields in vacuum. 
The two point sources at the grid/ABL interface radiate into both the grid 
and ABL. 
Since the two sources are uncorrelated with each other, their energy 
densities, instead of their field amplitudes, add in the grid. 
Thus no further modification of $\delta$ from that given by 
Eq.~\ref{eq:delta} is needed to satisfy Eq.~\ref{eq:FFTEME}.
It is numerically confirmed that $\delta$ does not depend on the total length 
of the system.

Examples of the noise source of electric field $E_s(t_j)$ at $T=30,000$ K 
and $T=50,000$ K are shown in Fig.~\ref{fig:fig1}. 
$\Delta x = 1$ nm, and $\Delta t \ll \tau_c$. 
The frequency range of interest is set as $\omega_{min}=2 \times 10^{15}$ Hz, 
$\omega_{max}=2.5 \times 10^{16}$ Hz, and the frequency resolution 
$\Delta \omega=1 \times 10^{12}$ Hz.
>From the condition $\Delta t < \pi/\omega_{max}$, $\Delta x$ shall be less 
than 37 nm.
Figure \ref{fig:fig2} compares the FDTD calculated energy density to that of 
thermal radiation density $D(\omega,T)$.
Using $\Delta x=10$ nm creates a slight discrepancy at high frequencies; 
e. g. at $\omega \leq 1 \times 10^{16}$ Hz the mean error $\gtrsim$ 2.5\%.
To reduce the error to below 2.5\% at $\omega_{max}=2.5 \times 10^{16}$ Hz, 
we refine the resolution. 
Using $\Delta x=4$ nm changes the error at $\omega_{max}$ to 1.6\%.
If the total time step $2M=2^{21}$ is fixed, the decrease of $\Delta t$ leads 
to a reduction of $\tau_{sim} = 2M \Delta t$, which increases 
$2 \pi / \tau_{sim}$ to $2 \times 10^{11}$ Hz. 
We must check that $2 \pi / \tau_{sim} < \omega_{min}$ and 
$2 \pi / \tau_{sim} < \Delta \omega$ are still satisfied. 
With $\Delta x=1$ nm, the error at $2.5 \times 10^{16}$ Hz is further reduced 
to $< 0.1$\%.
$2 \pi / \tau_{sim}$ increases to $9 \times 10^{11}$ Hz, 
which is still below the set values of $\omega_{min}$ and $\Delta \omega$. 
Therefore, using the value of $\delta$ in Eq.~\ref{eq:delta} and carefully 
choosing the numerical resolutions yield the blackbody spectrum at every 
point in the grid within the frequency range of interest.

Figure \ref{fig:fig3} compares the FDTD calculated temporal correlation 
function of the electric field to that of Eq.~\ref{eq:coran} at $T=30,000$ K 
and $50,000$ K.
With increasing temperature, the coherence time $\tau_c$ reduces quickly. 
The quantitative dependence of $\tau_c$ on $T$ is found to be 
$\tau_c \approx 1.32\hbar/kT$.
This $1/T$ dependence does not change for a dimensionality higher than one; 
only the prefactor changes \cite{kano62}. 
As the correlation time $\tau_c$ decreases, the time step $\Delta t$ shall 
be reduced to maintain the temporal resolution of the correlation function. 
The subsequent reduction of total running time $\tau_{sim} = 2 M \Delta t$ 
does not affect the numerical accuracy, as long as the total number of time 
steps $2M$ is fixed. 
A decrease of $2M$ would result in an increased mean-square error in the 
correlation function due to less sampling. 
As shown in Fig.~\ref{fig:fig3}, the good agreement of the FDTD calculated 
temporal correlation function to that of blackbody radiation given by 
Eq.~\ref{eq:bbu} confirms that introducing noise sources with the 
characteristics of blackbody radiation at the FDTD absorbing boundary 
effectively simulates thermal noise in vacuum.

\subsection{Thermal noise in a 1D cavity\label{subsec:1dcav}}

Next we calculate the thermal noise in a dielectric slab of length $L$ and 
refractive index $n > 1$. 
This slab constitutes an open cavity in that electromagnetic field leakage 
occurs from both surfaces of the slab into an exterior region. 
A schematic of the 1D open cavity is shown in the inset of Fig. 4(b). 
The cavity mode frequency is $\omega_m= m \pi c /nL$, where $m$ is an integer 
and $c$ is the speed of light in vacuum. 
The frequency spacing of adjacent modes is 
$d\omega =\omega_{m}-\omega_{m-1}=\pi c /nL$, which is independent of $m$. 
Ignoring intracavity absorption, the decay of cavity photons is caused only 
by their escape from the cavity.  
All the cavity modes have roughly the same decay time $\tau = 1/k_ic$, where 
$k_i= -\ln\left(r^2\right)/2nL$, and $r=(1-n)/(1+n)$ is the reflection 
coefficient at the boundary of the dielectric slab. 
The mode linewidth is $\delta\omega = 2/\tau$. 
We simulate only good cavities whose modes are well separated in frequency, 
namely, $\delta\omega < d \omega$. 
Since $\delta \omega \propto 1/L$, the ratio $\delta \omega / d\omega $ is 
independent of $L$, and is only a function of $n$.  

The Langevin equation for the annihilation operator $\hat{a}_m(t)$ of photons 
in the $m$-th cavity mode is
\begin{equation}
  \frac{d\hat{a}_m(t)}{dt} = -\frac{1}{\tau} \hat{a}_m(t) + \hat{F}_m(t),
  \label{langevin}
\end{equation}
where $\hat{F}_m(t)$ is the Langevin force.
If the noise correlation time $\tau_c \ll \tau$, $\hat{F}_m(t)$ can be 
considered $\delta$-correlated in time. 
The Markovian approximation gives  
$\left< \hat{F}_m^{\dagger}(t)\hat{F}_m(t')\right>=D_F\delta(t-t')$. 
According to the fluctuation dissipation theorem, 
$D_F=(1/\tau) n_T(\omega_m)$, where 
$n_T(\omega_m)=(e^{\hbar\omega_m/kT}-1)^{-1}$ is the number of thermal 
photons in a vacuum mode of frequency $\omega_m$ at temperature $T$ 
\cite{haken83}. 

From Eq. \ref{langevin}, the average photon number in one cavity mode 
$\left<\hat{n}_m(t)\right>\equiv 
\left<\hat{a}_m^{\dagger}(t)\hat{a}_m(t)\right>$ 
satisfies
\begin{equation}
  \frac{d}{dt}\left<\hat{n}_m(t)\right> = -\frac{2}{\tau} 
  \left<\hat{n}_m(t)\right> + \frac{2}{\tau} n_T(\omega_m).\label{eq:ada-de}
\end{equation}
At steady state, $\left< \hat{n}_m \right>= n_T(\omega_m)$ in
each cavity mode.
The number of thermal photons is determined by the Bose-Einstein distribution 
$n_T(\omega_m)$. 
$\left< \hat{n}_m \right>$ is independent of the cavity mode 
decay rate because the amount of thermal fluctuation entering the cavity 
increases by the same amount as the intracavity energy decay rate.

\begin{figure}
  \includegraphics[width=8.6cm]{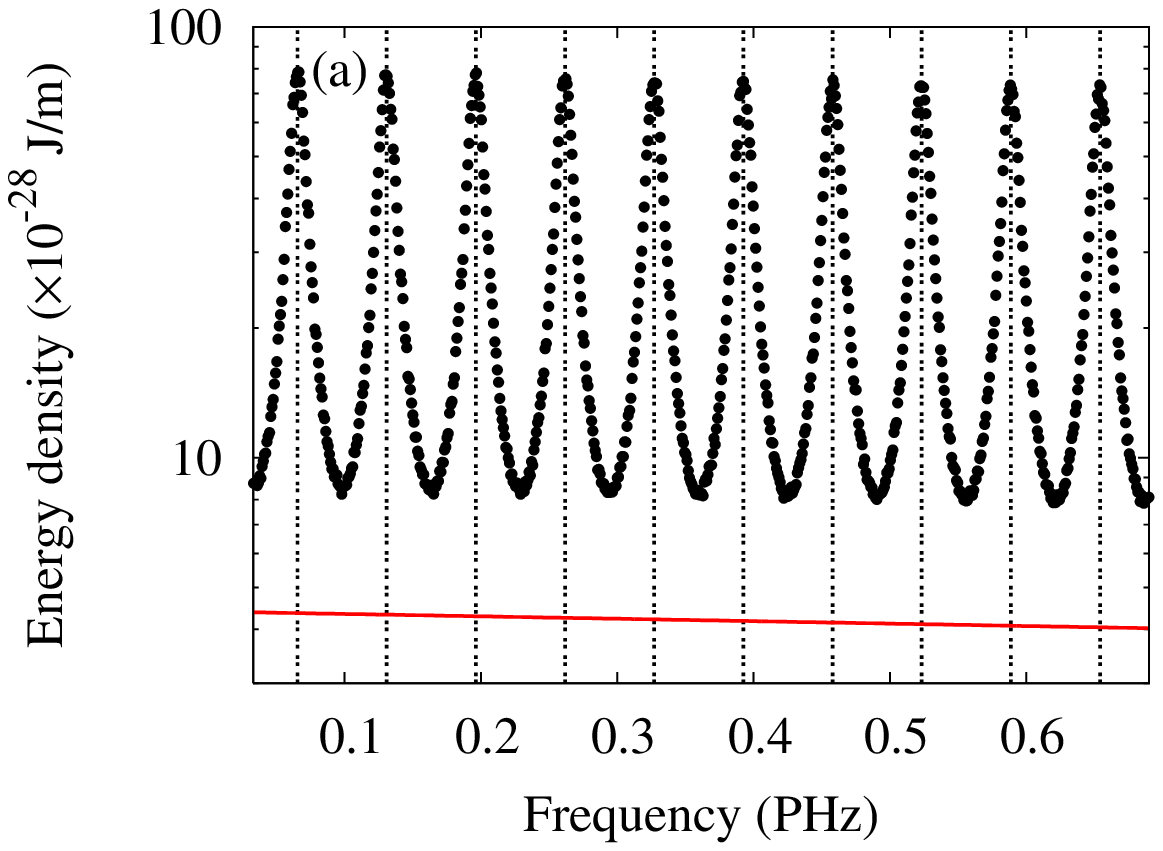}
  \includegraphics[width=8.6cm]{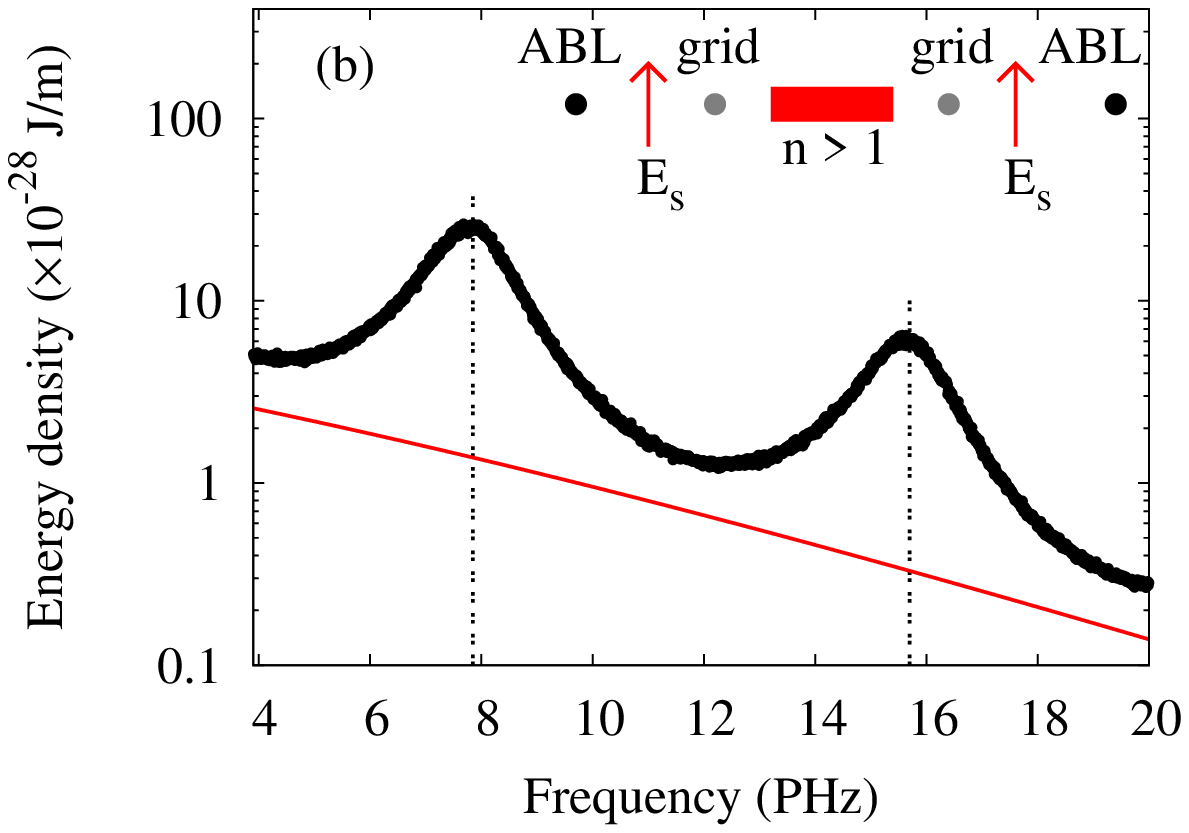}
  \caption{\label{fig:fig4} (Color online) 
    Spatially-averaged EM energy density $U(\omega)$, calculated by
    the FDTD method, versus frequency $\omega$ in a dielectric slab cavity
    with refractive index $n = 6$ and length $L = 2400$ nm (a) and
    $L = 20$ nm (b).
    The vertical black dashed lines mark the frequencies of cavity modes
    $\omega_m$.
    The spectrum of impinging blackbody radiation $D(\omega,T)$ is also
    plotted (red solid line).
    In (a) the cavity decay time $\tau=143$ ps, much longer than $\tau_c$.
    In (b), $\tau=1.19$ fs, comparable to $\tau_c$.
  }
\end{figure}

\begin{figure}
  \includegraphics[width=8.6cm]{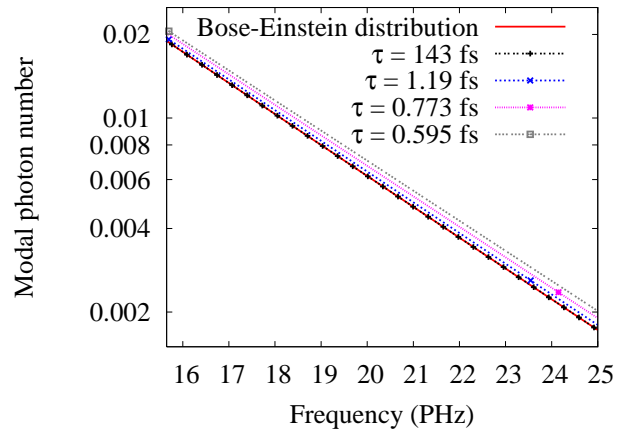}
  \caption{\label{fig:fig5} 
    (Color online) 
    The number of thermal photons in individual cavity modes $n_m$, 
    calculated via the FDTD method, for a slab cavity with $n=6$.
    The cavity length $L$ is varied to change $\tau$.
    The impinging blackbody radiation has $T=30,000$ K and $\tau_c=0.337$ fs.
    The values of $\tau_c/\tau$ are 0.0024, 0.29, 0.43, and 0.56.
    Lines are drawn to connect the data points at the mode frequencies
    $\omega_m=\pi c m/nL$ to illustrate its frequency dependence.
    For $\tau_c\ll\tau$ (only every 5th mode for $\tau=143$ fs is shown to
    improve the visibility), the photon number $n_m$ coincides with the
    Bose-Einstein distribution $n_T$.
    When $\tau_c \sim \tau$, $n_m$ deviates from $n_T$.
  }
\end{figure}

In our FDTD simulations, we verify that when $\tau \gg \tau_c$ the number of 
thermal photons in one cavity mode is equal to $n_T(\omega_m)$. 
Since there is neither a driving field (e.g. a pumping field) nor excited 
atoms in the cavity, the EM energy stored in one cavity mode comes entirely 
from the thermal radiation of the ABL which is coupled into that particular 
mode.
The steady-state number of photons in the $m$-th cavity mode is obtained from 
the FDTD calculation of intracavity EM energy within the frequency range 
$\omega_{m-1/2} < \omega < \omega_{m+1/2}$, where 
$\omega_{m \pm 1/2} = (m \pm 1/2) \pi c /nL$.  
\begin{align}
  &n_m \equiv \left<\hat{n}_m \right>=\frac{1}{\hbar\omega_m} 
  \int_{\omega_{m-1/2}}^{\omega_{m+1/2}}d\omega \nonumber\\
  &\times\int_0^L dx
  \left(\frac{1}{2}\epsilon|E(x,\omega)|^2+\frac{1}{2}\mu_0|H(x,\omega)|^2 
  \right) \label{eq:noisem}
\end{align}
In our simulation, the temperature of the thermal sources at the ABL is 
$T=30,000$ K. 
The coherence time of thermal radiation is $\tau_c=0.337$ fs. 
The refractive index of the dielectric slab is $n=6$, and the length is 
$L=2400$ nm. 
The cavity lifetime $\tau=143$ fs, is much longer than  $\tau_c$. 
The reason to choose a relatively large value of $n$ is to have 
$\delta \omega < d \omega$ so that the cavity modes are separated in 
frequency.  
Care must be taken in setting the grid resolution because the intracavity 
wavelength is reduced to $\lambda/n$. 
To maintain the spatial resolution, $\Delta x$ is decreased to meet 
$\Delta x \ll \lambda/n$. 
In our FDTD simulation, $\Delta x=1$ nm and $2M=2^{21}$. 
After the intracavity EM field reaches the steady state, we calculate the 
average thermal energy density inside the cavity 
\begin{equation}
  U(\omega) = \frac{1}{L}\int_0^L dx \left(\frac{1}{2}\epsilon|E(x,\omega)|^2+
  \frac{1}{2}\mu_0|H(x,\omega)|^2 \right).
\end{equation}
Figure \ref{fig:fig4}(a) shows the intracavity noise spectrum $U(\omega)$, 
which features peaks at the cavity resonant frequencies $\omega_m$.
Because $n>1$, EM energy is also stored in the dielectric slab at 
frequencies away from cavity resonances. 
For example, $U(\omega=\omega_{m\pm 1/2})$ is higher than that in vacuum by a 
factor of $2n^2/(n^2+1)$. 
Thus the entire noise spectrum lies on top of the vacuum blackbody radiation 
spectrum, as confirmed in Fig.~\ref{fig:fig4}.
The number of thermal photons in a cavity mode is calculated via 
Eq.~\ref{eq:noisem} and plotted in Fig.~\ref{fig:fig5}.
The modal photon number $n_m$ is equal to $n_T(\omega_m)$ with a mean error 
less than 0.1\%. 
This result agrees with the steady-state solution of Eq.~\ref{eq:noisem}. 
It confirms that our numerical model of thermal noise in an open cavity is 
consistent with the prediction of quantum mechanical theory.
Note that the modal photon numbers in Fig. 5 are time averaged values. 
Their values being much less than unity can be interpreted in a quantum 
mechanical picture as that most of the time there is no photon in the cavity 
mode.

The above calculation is done in the Markovian regime. 
Next we move to the non-Markovian regime by reducing $\tau$. 
The refractive index is kept at $n=6$, while the cavity length $L$ is 
reduced. 
This is a simple way of increasing the mode linewidth $\delta \omega$ while 
keeping the modes separated in frequency, i.e. keeping 
$\delta\omega / d\omega$ constant. 
If $\tau$ is decreased to less than $\tau_c$, $\Delta t$ shall be reduced to 
keep $\Delta t \ll \tau$. 
Meanwhile, the increase of the mode linewidth and mode spacing allows low 
frequency resolution, namely, an increase of $\Delta \omega$. 
For example, at $L =  20$ nm, we set $\Delta x=0.1$ nm, 
$\Delta\omega=9 \times 10^{12}$ Hz and $2M=2^{21}$. 
Figure \ref{fig:fig4}(b) shows the intracavity noise spectrum $U(\omega)$ 
in this regime, which also features peaks at the cavity resonant frequency 
$\omega_m$.
Figure~\ref{fig:fig5} shows the FDTD-calculated value of $n_m$ as $L$ 
decreases gradually from $2400$ nm to $20$ nm. 
When $\tau$ approaches $\tau_c$, $n_m$ is no longer independent of $\tau$, 
but  starts increasing from $n_T(\omega_m)$. 
This means the number of thermal photons that are captured by a cavity mode 
increases with the decrease of $\tau$.
As the coherence time of thermal radiation impinging onto the cavity 
approaches the cavity photon lifetime, the constructive interference of the 
thermal field is improved inside the cavity, leading to a larger buildup of 
intracavity energy.

\begin{figure}
  \includegraphics[width=8.6cm]{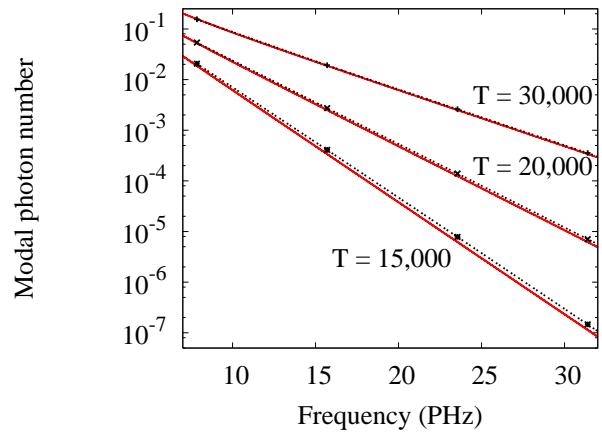}
  \caption{\label{fig:fig6} (Color online)
    The number of thermal photons in individual cavity modes $n_m$,
    calculated via the FDTD method, for a dielectric slab cavity with $n=6$
    and $L= 20$ nm.
    The cavity decay time $\tau=1.19$ fs.
    The temperature of blackbody radiation is varied to change $\tau_c$.
    The values of $\tau_c/\tau$ are 0.29 ($T=30,000$ K), 0.43 ($T=20,000$ K),
    and 0.56 ($T=15,000$ K).
    Black dashed lines are drawn to connect the data points at the mode
    frequencies $\omega_m=\pi c m/nL$ to illustrate its frequency dependence.
    For comparison, the Bose-Einstein distribution $n_T(\omega)$ is also
    plotted (red solid lines) for the same temperatures.
  }
\end{figure}

We also investigate a different situation where $\tau$ is fixed and $\tau_c$ 
is varied.
By reducing the temperature $T$, the coherence time of thermal radiation 
$\tau_c$ is increased.  
Meanwhile, the energy density of thermal radiation is decreased. 
As shown in Fig.~\ref{fig:fig6}, the number of thermal photons $n_m$ in a 
cavity mode is reduced. 
This can be easily understood as there are fewer thermal photons incident 
onto the cavity at lower $T$. 
Nevertheless, $n_m$ is larger than $n_T(\omega_m)$ at the same $T$. 
This is because the longer coherence time of the thermal field results in 
better constructive interference inside the cavity.

\section{Discussion\label{sec:discussion}}

To gain a better understanding of our FDTD simulation results in the 
non-Markovian regime, we analytically examine the effect of noise 
correlation time $\tau_c$ on the amount of thermal noise inside an open 
cavity. 
The ratio of the intracavity EM energy at frequency $\omega$ to the energy 
density of the thermal source outside the cavity is
$W(\omega) \equiv \left\{\int_0^L dx \left[\frac{1}{2} 
\epsilon|E(x,\omega)|^2+ 
\frac{1}{2} \mu_0|H(x,\omega)|^2\right]\right\}/D(\omega,T)$.
For a dielectric slab of refractive index $n$ and length $L$, we obtain the 
expression for $W(\omega)$ using the transfer matrix method \cite{born75}, 
\begin{equation}
  W(\omega) = \frac{2 nc}{\omega}\left[\frac{2\omega nL(1+n^2)/c + (n^2-1)
      \sin(2\omega nL/c)}
    {1+6n^2+n^4-(n^2-1)^2\cos(2\omega nL/c)}\right]
\end{equation}
It can be used to calculate the ratio 
$B_m(\tau_c,\tau) \equiv  {n}_m / n_T(\omega_m)$, as 
\begin{equation}
  n_m= \left[ \int_{\omega_{m-1/2}}^{\omega_{m+1/2}} d\omega \, W(\omega) 
    D(\omega, T) \right] / \hbar \omega_m.
  \label{n_m}
\end{equation}

\begin{figure}
  \includegraphics[width=8.6cm]{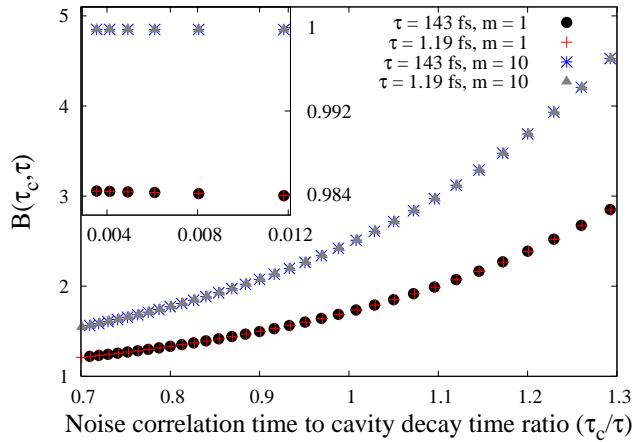}
  \caption{\label{fig:fig7} (Color online)
    $B_m(\tau_c,\tau) \equiv n_m / n_T(\omega_m)$ vs. the ratio of noise
    correlation time to cavity decay time $\tau_c/\tau$.
    $B_m(\tau_c,\tau)$ depends only on the ratio $\tau_c/\tau$ and not on
    $\tau_c$ nor $\tau$ individually.
    This plot was generated by fixing $\tau$ and varying $\tau_c$.
    Cavity modes of higher $m$ have a larger value of $B_m(\tau_c,\tau)$
    whether in the regime $\tau_c\sim\tau$ (main panel) or $\tau_c\ll\tau$
    (inset).
  }
\end{figure}

In the Markovian regime $\tau_c \ll \tau$ and $D(\omega,T)$ is nearly 
constant over the frequency interval of one cavity mode so
\begin{align}
  B_m(\tau_c,\tau) &= \frac{D(\omega_m,T)}{\hbar \omega_m n_T(\omega_m)}
  \int_{\omega_{m-1/2}}^{\omega_{m+1/2}} d\omega \, W(\omega) \nonumber\\  
  &= \frac{1}{\pi c}\int_{\omega_{m-1/2}}^{\omega_{m+1/2}} d\omega \, 
  W(\omega).
\end{align}
We input the same parameters as the FDTD simulation: $n=6$, $L=2400$ nm, 
and $\tau=143$ fs.  
As $\tau_c$ approaches zero, the integration of $W(\omega)$ from 
$\omega_{m-1/2}$ to $\omega_{m+1/2}$ gives a value close to $\pi c$. 
Thus, as shown in the inset of Fig.~\ref{fig:fig7}, 
$B_m(\tau_c,\tau)\approx 1$ for $\tau_c/\tau\ll 1$.
The deviation of $B_m(\tau_c,\tau)$ from one is greater for smaller $m$. 
One possible reason is that the condition $\delta\omega\ll \omega_m$ no 
longer holds for small $m$ and there is a large uncertainty in defining the 
frequency of a cavity mode whose linewidth is comparable to its center 
frequency. 
In other words, the calculation of $n_m$ using Eq.~\ref{n_m} becomes 
questionable.

In the non-Markovian regime $\tau_c\gtrsim \tau$, $D(\omega,T)$ is not 
constant over the frequency range of a cavity mode anymore, thus it cannot be 
taken out of the integral in Eq.~\ref{n_m}. 
The behavior of $B_m(\tau_c,\tau)$ in this regime is shown in the main panel 
of Fig.~\ref{fig:fig7}. 
As $\tau_c$ approaches $\tau$, $B_m(\tau_c,\tau)$ no longer stays near one 
but increases with $\tau_c/\tau$. 
This result is consistent with that of the FDTD simulation presented in the 
previous section. 
On one hand, if $\tau$ is fixed and $\tau_c$ is increased by decreasing the 
temperature $T$, the absolute number of thermal photons in a cavity mode 
$n_m$ decreases, but its ratio to the number of thermal photons in a vacuum 
mode $n_T(\omega_m)$ increases. 
On the other hand, if $\tau_c$ is fixed and $\tau$ is decreased by shortening 
the cavity length $L$, both $n_m$ and $n_m / n_T(\omega_m)$ increase. 
The departure of ${n}_m$ from $n_T(\omega_m)$ is a direct consequence of the 
breakdown of the Markovian approximation. 
When the coherence time of thermal radiation is comparable to the cavity 
decay time, the Langevin force $\hat{F}_m(t)$ in Eq.~\ref{langevin} is no 
longer $\delta$-correlated in time, and Eq.~\ref{eq:ada-de} is invalid. 

\section{Conclusion\label{sec:conclusion}}

We have calculated the fluctuations of EM fields in open cavities due to 
output coupling with the FDTD method. 
The fluctuation dissipation theorem dictates that the cavity field 
dissipation by leakage be accompanied by thermal noise, which is simulated 
here by classical electrodynamics.  
The absorbing boundary of the FDTD grid is treated as a blackbody, which 
radiates into the grid. 
We have synthesized the noise sources whose spectrum is equal to that of 
blackbody radiation. 
Careful selection of numerical parameters in the FDTD simulation avoids the 
distortion of the noise spectrum by wave propagation in the 1D grid. 
It is numerically confirmed that the noise fields propagating in vacuum 
retain the blackbody spectrum and temporal correlation function. 
When a cavity is placed in the grid, the thermal radiation is coupled into 
the cavity and contributes to the thermal noise for the cavity field. 
We calculate the thermal noise in a 1D dielectric slab cavity. 
In the Markovian regime where the cavity photon lifetime $\tau$ is much 
longer than the coherence time of thermal radiation $\tau_c$, the 
FDTD-calculated amount of thermal noise in a cavity mode agrees with that 
given by the quantum Langevin equation. 
This validates our numerical model of thermal noise which originates from 
cavity openness or output coupling. 
Our FDTD simulation also demonstrates that in the non-Markovian regime the 
steady-state number of thermal photons in a cavity mode exceeds that in a 
vacuum mode. 
This is attributed to the constructive interference of the thermal field 
inside the cavity. 

The advantage of our numerical model is that the thermal noise is added in 
the time domain without the knowledge of cavity modes. 
It can be applied to simulate complex open systems whose modes are not known 
prior to the FDTD calculations. 
Our approach is especially useful for very leaky cavities whose modes 
overlap strongly in frequency, as the thermal noise related to the cavity 
leakage is introduced naturally without distinguishing the modes. 
Therefore, we believe the method developed here can be applied to a whole 
range of quantum optics problems.  
Although in this paper the FDTD calculation of thermal noise is performed on 
1D systems, the extension to 2D and 3D systems is straightforward. 
We comment that our approach does not apply to the simulation of zero-point 
fluctuation which has a different physical origin from the thermal noise.
However, our numerical method can be used to study the dynamics of EM
fields which are excited by arbitrarily correlated noise sources.
One potential application is noise radar \cite{horton,walton}.
The propagation, reflection and scattering of ultra-wideband signals utilized
by noise radar can be easily simulated using the method developed here.

\end{document}